\documentstyle[aps,multicol,prl,epsf]{revtex}
\begin{document}
\draft
\title{Which phase is measured in the mesoscopic Aharonov-Bohm interferometer?}
\author{A. Aharony$^a$, O. Entin-Wohlman$^a$, B. I. Halperin$^b$ and Y. Imry$^c$}
\address{$^a$School of Physics and Astronomy, Raymond and Beverly Sackler
Faculty of Exact Sciences, \\ Tel Aviv University, Tel Aviv 69978,
Israel\\ }
\address{$^b$Department of Physics, Harvard University, Cambridge, MA 02138}
\address{$^c$Department of Condensed Matter Physics, The Weizmann
Institute of Science, Rehovot 76100, Israel}

\date{\today}
\maketitle
\begin{abstract}
Mesoscopic solid state Aharonov-Bohm interferometers have been
used to measure the ``intrinsic" phase, $\alpha_{QD}$, of the
resonant quantum transmission amplitude through a quantum dot
(QD). For a two-terminal ``closed" interferometer, which conserves
the electron current, Onsager's relations require that the
measured phase shift $\beta$ only ``jumps" between 0 and $\pi$.
Additional terminals open the interferometer but then $\beta$
depends on the details of the opening. Using a theoretical model,
we present quantitative criteria (which can be tested
experimentally) for $\beta$ to be equal to the desired
$\alpha_{QD}$: the ``lossy" channels near the QD should have both
a small transmission and a small reflection.
\end{abstract}

\begin{multicols}{2}

\section{Introduction}

Recent advances in the fabrication of nanometer scale electronic
devices raised much interest in the quantum mechanics of quantum
dots (QDs), which represent artificial atoms with experimentally
controllable properties\cite{review,book}. The quantum nature of
the QD is reflected by resonant tunneling through it, as measured
when the QD is connected via metallic leads to electron
reservoirs. The measured conductance $G$ shows peaks whenever the
Fermi energy of the electrons crosses a bound state on the
QD\cite{note}. Experimentally, the energies of these bound states
are varied by controlling the plunger gate voltage on the QD, $V$.
Quantum mechanically, the information on the tunneling is
contained in the complex transmission amplitude,
$t_{QD}=\sqrt{T_{QD}}e^{i\alpha_{QD}}$. It is thus of great
interest to measure both the magnitude $T_{QD}$ and the phase
$\alpha_{QD}$, and study their dependence on $V$\cite{englman}.
Although the former can be deduced from measuring $G$, via the
Landauer formula \cite{landauer}, $G=\frac{2e^2}{h}T$,
experimental information on the latter has only become accessible
since 1995\cite{yacoby,schuster}, using the Aharonov-Bohm (AB)
interferometer\cite{azbel}.

\begin{figure}
\leftline{\epsfclipon\epsfxsize=3.in\epsfysize=5in\epsffile{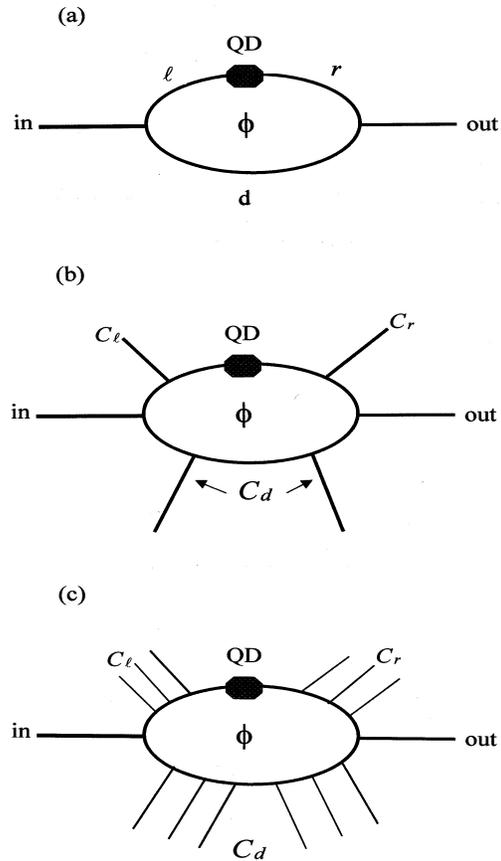}}
\vskip .4truecm \caption{Model for the AB interferometer: (a)
Closed two-terminal case, (b) schematic picture of the
six-terminal open interferometer, (c) model for the open
interferometer.} \label{AB}
\end{figure}

In the AB interferometer, an incoming electronic waveguide is
split into two branches, which join again into the outgoing
waveguide (see Fig. \ref{AB}(a)). Aharonov and Bohm\cite{AB}
predicted that a magnetic flux $\Phi$  through the ring would add
a difference $\phi=e\Phi/\hbar c$ between the phases of the wave
functions in the two branches of the ring, yielding a periodic
dependence of the overall transmission $T$ on $\phi$. Placing a QD
on one of the branches, one expects $T$ also to depend on
$t_{QD}$. Indeed, the experiments found a periodic dependence of
$T(\phi)$, and fitted the results to a Fourier expansion of the
form
\begin{equation}
T=A+B\cos(\phi+\beta)+C\cos(2 \phi+\gamma)+\ldots, \label{fit}
\end{equation}
with the conventions $B,~C>0$.

 In a simple two-slit situation, there is no reflection of
 electrons from either the source or the ``screen" which collects
 them. Therefore,
the electron passes through each branch (including the QD) only
once, and the total transmission amplitude is equal to the sum of
the {\it amplitudes} in the two branches,
\begin{equation}
t=t_1 e^{i\phi}+t_2. \label{2slit}
\end{equation}
(Gauge invariance allows one to attach the AB phase $\phi$ to
either branch). Assuming also that $t_1=|t_1|e^{i\alpha_1}=c
t_{QD}$, and that both $c=|c|e^{i\delta}$ and
$t_2=|t_2|e^{i\alpha_2}$ do not depend on $V$ near the QD's
resonances, one obtains Eq. (\ref{fit}), with $B=2|ct_2t_{QD}|$,
$C=0$ (i.e. no higher harmonics) and
$\beta=\alpha_{QD}+\delta-\alpha_2$. Below we subtract from
$\alpha_{QD}$ and from $\beta$ their values at large negative $V$,
far away from the resonances, thus removing $V$-independent
quantities like $\delta-\alpha_2$. For the ``closed" two-terminal
geometry of Fig. \ref{AB}(a), as used by Yacoby {\it et
al.}\cite{yacoby}, the expectation that $\beta=\alpha_{QD}$
(equivalent to the two-slit situation)  was clearly not borne out
by the measurements: Unitarity (conservation of current) and time
reversal symmetry imply the Onsager relations \cite{onsager,but},
which state that $G(\phi)=G(-\phi)$, and therefore $\beta$ (as
well as $\gamma$ etc.) {\it must} be equal to zero or $\pi$.
Indeed, the experimental \cite{yacoby} $\beta$ ``jumps" from 0 to
$\pi$ whenever $V$ crossed a resonance of the QD, and then
exhibits an a priori unexpected ``phase lapse" back to 0, between
every pair of resonances. Later experiments\cite{schuster} opened
the interferometer, using the six-terminal configuration shown
schematically in Fig. \ref{AB}(b); the additional leads allow
losses of electronic current, thus breaking unitarity. Indeed, the
resulting data gave a gradual increase of $\beta$ through each
resonance, accompanied by a peak in the amplitude $B$, but
maintained the sharp ``phase lapse" back to zero between
resonances, which were accompanied by zeroes in $B$. In the
present paper we present a theoretical model, aimed to imitate the
experimental setups of Fig. \ref{AB}(a) and (b). Figure \ref{ziur}
shows examples of our model calculations for $A,~B,~C$ and $\beta$
versus $V$. Qualitatively, these plots look similar to those found
experimentally\cite{yacoby,schuster}. However, as discussed below,
the quantitative results for the open interferometers depend on
details of the opening.

The above experimental results led to much theoretical discussion.
Some of this \cite{wu,kang} emphasized the non-trivial effects of
the ring itself on the measured results, even for the closed case.
Other theoretical papers
\cite{hack,oreg,ryu,xu,lee,silvestrov,levy} {\it assumed} that the
measured $\beta$ represents the correct $\alpha_{QD}$, and
discussed the possible origins of the observed features, e.g. the
``phase lapse" and the similarity between the data at many
resonances. However, not much attention was given to the {\it
validity of this assumption}. Since $\beta$ is equal to 0 or $\pi$
for the closed interferometer, and deviates from these values for
the open one, it is clear that $\beta$ {\it must} depend on the
details of {\it how the system was opened}. Indeed, Ref.
\cite{prl} considered one example of an open interferometer, and
showed that the deviation of $\beta$ from its trivial values (0
and $\pi$) increases monotonically with the strength of the
coupling to the ``lossy" channel. Although different values of
this coupling gave {\it qualitatively} similar $\beta(V)$ curves,
which were also similar to the experimental results, the detailed
dependence of $\beta$ on $V$ varied with that strength. As a
result, Ref. \cite{prl} posed the challenge of finding clear
criteria as to when the experimental $\beta$ is really equal to
the intrinsic $\alpha_{QD}$.

\begin{figure}
\leftline{\epsfclipon\epsfxsize=2.5in\epsfysize=3in\epsffile{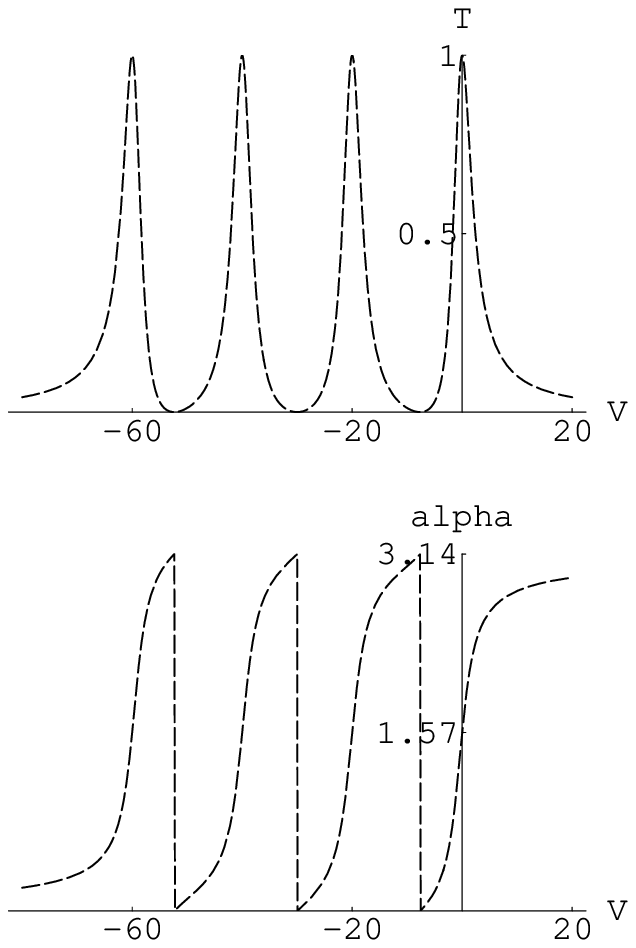}}
\vskip .4truecm \caption{Transmission $T_{QD}$ and ``intrinsic"
phase $\alpha_{QD}$ for $N=4$ states on the QD, with ``gap"
$U=20J$, versus the gate voltage $V$ (in units of $J$).}
\label{alpha}
\end{figure}

In the present paper we address this challenge\cite{weiden}. Sec.
II presents a simple model for the QD, which contains resonances
and ``phase lapses". Typical results for the ``intrinsic" $T_{QD}$
and $\alpha_{QD}$ are shown in Fig. \ref{alpha}. The latter is
also reproduced in Fig. \ref{ziur} (calculated with the same QD
parameters), for comparison with $\beta$. We are not aware of any
earlier quantitative comparisons of this kind. Sections III and IV
then present a simple model for the (closed and open)
interferometer, and discusses the optimal way to open the
interferometer, so that the ``measured" $\beta$ will be close to
the theoretical ``intrinsic" $\alpha_{QD}$. Our exact analytical
results confirm the intuitive expectations of Ref. \cite{prl}: to
have $\beta=\alpha_{QD}$, the electron must cross each branch only
once.  One necessary condition for this was appreciated
qualitatively before \cite{prl}:  the electron must practically
never be reflected from the ``forks" where the ring meets the
incoming and outgoing terminals, in order to recover the two-slit
result (\ref{2slit}). In our model, this is achieved by having a
very small net transmission after crossing each of the additional
``lossy" channels $C_\ell,~C_r$ and $C_d$ in Fig. \ref{AB}(b).
However, we find two additional conditions: first, the
transmission through the upper branch, $t_1$, should have the same
phase (up to a $V$-independent additive constant) as $t_{QD}$,
i.e. $\alpha_{QD}$. In general, the scattering of the electron
from the gates into channels $C_\ell$ and $C_r$ might cause
``rattling" of the electron back and forth through the QD,
introducing more phase shifts into $t_1$. We avoid that by also
having a very {\it small reflection} from the scatterers $C_\ell$
and $C_r$. Below we introduce a parameter, $J_x$, which relates to
the tunneling probabilities of the electron from the ring onto the
``lossy" channels. As $J_x$ increases, the transmission through
the ``lossy" scatterers decreases, but the reflection from them
increases. Therefore, there is only an {\it intermediate} range of
$J_x$ where $\beta=\alpha_{QD}$ (shown in the lower left box in
Fig. \ref{ziur}). The second new condition is that there should be
no direct losses from the QD itself; as explained below, these
``smear" the ``phase lapses". In Sec. V we discuss these results,
and propose additional experiments which would check if an open
interferometer indeed reproduces the desired ``intrinsic" QD
information.

\section{Model for the QD}

As in many earlier
calculations\cite{wu,levy,hartzstein,damato,koval}, our analytic
calculations are based on the single-electron tight-binding model
(which can be viewed as a finite difference version of the
continuum case): the system is made of discrete sites $\{ i \}$,
with nearest neighbor (nn) real tunneling amplitudes $-J_{ij}$ and
site energies $\epsilon_i$. All nn distances are set equal to $a$.
The Schr\"odinger wave equation is thus written as
$(E-\epsilon_i)\psi_i=-\sum_jJ_{ij}\psi_j$, where the sum is over
nn's of $i$. In each calculation, we have a scattering element
connected to two one-dimensional (1D) leads, which have
$J_{i,i+1}=J,~\epsilon_i=0$. The scattering solution for a wave
coming from the left, with wave vector $k$ and energy $E=-2J\cos
ka$, is described by $\psi^L_m=e^{ikam}+re^{-ikam}$ on the left,
and by $\psi^R_m=t e^{ikam}$ on the right. The calculation of the
transmission and reflection amplitudes, $t$ and $r$, then amounts
to solving a finite set of linear equations for the wave functions
inside the scatterer.

The QD may be described as a single dot, with many discrete energy
levels. We model it by a set of smaller dots, each containing a
single resonant state, with energy
$\{E_R=\epsilon_{QD}=E_R(n),~n=1,...,N \}$. Each such state is
connected to its left and right nn's on the leads via bonds with
hopping amplitudes $\{ -J_L(n),~-J_R(n),~n=1,...,N \}$. The QD can
thus be described by $N$ wave functions $\psi_n$, obeying
$[E-E_R(n)]\psi_n=-J_L(n)\psi^L_0-J_R(n)\psi^R_0$ (where we choose
$\psi^L_0=1+r,~\psi^R_0=t$). The exact transmission amplitude is
easily found to be
\begin{equation}
t_{QD}=\frac{S_{LR} 2 i \sin ka}
{(S_{LL}+e^{-ika})(S_{RR}+e^{-ika})-|S_{LR}|^2}, \label{tintr}
\end{equation}
where $S_{XY}=\sum_{n}J_{X}(n)J_Y(n)^\ast/[E-E_R(n)]/J,~X,Y=L,R$
represent ``bare" Green's functions for sites $L$ and $R$.

In the following, we use equidistant bound state energies,
$E_R(n)=V+U(n-1)$. The ``gap" $U$ can be viewed as the Hartree
energy for an electron added to a QD which already has $n-1$ other
electrons\cite{hack}, thus capturing some aspects of the {\bf
Coulomb blockade} behavior of the scattered electron. We study
$t_{QD}$ as function of the energy $V$, which represents the
plunger gate voltage on the QD. Fig. \ref{alpha} shows typical
results for the transmission $T_{QD}$ and for the ``intrinsic"
phase $\alpha_{QD}$, where the zero of $\alpha_{QD}$ is set at its
($k-$dependent) value at large negative $V$. In this figure and
below, we choose $ka=\pi/2$, so that $E=0$ and the resonances of
the transmission, where $T_{QD}=1$, occur exactly when
$E_R(n)=E=0$, i.e. when $V=-U(n-1)$ \cite{T0}. Results are not
sensitive to $k$ near the band center. We also use the simple
symmetric case, $J_L(n)=J_R(n) \equiv J$, and measure all energies
in units of $J$. Interestingly, this model reproduces the
apparently observed behavior of $\alpha_{QD}$: it grows smoothly
from 0 to $\pi$ as $E$ crosses $E_R(n)$, and exhibits a sharp
``phase lapse" from $\pi$ to 0 between neighboring resonances, at
points where $T_{QD}=0$. These latter points, associated with
zeroes of $S_{LR}$, represent Fano-like destructive interference
between the states on the QD \cite{fano,ryu,xu,sun,jlt}.

 Many earlier theoretical (e.g. \cite{hack}) and experimental
(e.g. \cite{schuster}) papers approximated $t$ by a sum of the
single resonance Breit-Wigner-like (BW) expressions\cite{BW},
\begin{eqnarray}
t \approx \sum_n \frac{e^{2ika}2 i \sin ka
J_L(n)J_R(n)^\ast}{E-E_R(n)+e^{ika}[|J_L(n)|^2+|J_R(n)|^2]/J}.
\label{BW}
\end{eqnarray}
Although this form gives an excellent approximation for $t_{QD}$
near each resonance, it completely misses the Fano-like zeroes and
the ``phase lapses" between resonances. This happens because the
approximation moves the zeroes off the real energy axis\cite{sun}.
As a result, the approximate $\alpha_{QD}$ never reaches 0 or
$\pi$, and exhibits a smooth decrease from a maximum to a minimum
near the correct ``phase lapse" values of $V$.  Since our aim here
is to check on accurate measurements of the ``intrinsic" phase,
for a broad range of the parameters, and since the phase lapse has
been a topic of much recent discussion
\cite{hack,oreg,ryu,xu,lee,silvestrov,levy}, we prefer to use the
exact solutions everywhere. This is particularly important since
typically, available experimental data \cite{schuster} show quite
broad resonances, so that the BW approximation is bound to fail
between them.

We emphasize again: in spite of the close similarity of our
``intrinsic" transmission results with the experiments, the
purpose of this paper is not to relate the calculated $t_{QD}$ to
the experimental systems. This would require a justification for
our choice of the same $J_L(n)$'s and $J_R(n)$'s for all the
resonances, which goes beyond the scope of the present paper.
Rather, we aim to check when the AB interferometer reproduces the
``input" behavior of the QD, by yielding $\beta=\alpha_{QD}$ for
all $V$. If this fails for our simple model then it would surely
fail in the more complicated cases, where electron-electron
interactions (beyond our simple Hartree approximation) become
important \cite{ji}.

\section{Model for the closed AB interferometer}

We next place the above QD on the upper branch of the closed AB
interferometer, as shown in Fig. \ref{AB}(a).  We now treat the
whole ring as our scatterer: each segment $s$ on the ring is
modeled by a 1D tight binding model of $M_s$ sites, with
$\epsilon_i=0$ and $J_{i,i+1}=J_{s}$ ($s=\ell,~r,~d$ for the left
and right upper segments and for the lower path, respectively).
Taking advantage of gauge invariance, we attach the AB phase
factor $e^{i\phi}$ to the hopping amplitude from the right hand
``fork" onto its nn on branch $r$, which we write as $-J_r
e^{i\phi}$. Writing the wave functions in segment $s$ as
$\psi^s_m=A_s \eta_s^m+B_s \eta_s^{-m}$, with $\eta_s$ given by
$E=-J_s(\eta_s+\eta_s^{-1})$, it is easy to express the total
transmission and reflection amplitudes through the interferometer,
$t$ and $r$, in terms of the six amplitudes $\{A_s,~B_s \}$, and
obtain six linear equations whose coefficients also contain $\{
S_{XY} \}$. Having solved these equations, one finally finds that
the dependence of the total transmission amplitude $t$ on the AB
phase $\phi$ has the general form
\begin{equation}
t=\frac{F+G e^{-i \phi}}{W+Z \cos \phi}, \label{tt}
\end{equation}
where the complex functions $F,~G,~W$ and $Z$ all depend on the
other parameters of the QD (including $V$), the interferometer and
the electron wave vector $k$. It is easy to convince
oneself\cite{prl} that, apart from an overall multiplicative
factor, the numerator represents the two-slit situation of one
crossing through each branch of the ring, while the $\cos\phi$
term in the denominator comes from a sum over an infinite
geometrical series of additional motions around the ring: clock-
and counterclockwise contributions contain factors of $e^{i\phi}$
and $e^{-i\phi}$, multiplying the same complex coefficient. Except
for the detailed dependence of the coefficients on $V$, these
facts are model independent. In fact, the form (\ref{tt}) appeared
in many earlier model calculations (e.g.
\cite{azbel,hartzstein,prl,weiden}). In fact, Eq. (\ref{tt})
implies that the exact form for $T(\phi)$ is
\begin{equation}
T(\phi)=|t|^2=\frac{A+B\cos(\phi+\beta)}{1+P\cos\phi+Q\cos^2\phi}.\label{TT}
\end{equation}
A fit to this equation, instead of the Fourier expansion
(\ref{fit}), would be much more accurate (with only five
parameters), and would enable an easier comparison of the data
with theoretical calculations of $F,~G,~W$ and $Z$.

Using exact integration on $T(\phi)$, the $V$-dependence of the
coefficients in the expansion (\ref{fit}) are presented for the
closed interferometer in Fig. \ref{ziur} (upper left), for the
same QD parameters as in Fig. \ref{alpha}. (The figure was
produced with $M_\ell=M_r=6,~M_d=12$, but the results for the
closed case do not depend on these numbers).  For the closed
interferometer, time reversal symmetry implies that the ratio
$F/G$ in Eq. (\ref{tt}) must be real, and thus $T$ depends only on
$\cos \phi$, in agreement with Onsager's relations. This yields
the same jumps of $\beta$ between zero and $\pi$ as in Yacoby {\it
et al.}'s experiments\cite{yacoby}, coincident with peaks and
zeroes of $B$.

\section{Model for the open AB interferometer}

Pursuing one possible scenario \cite{prl}, we model the ``leaking"
from each of the three segments on the ring (imitating
$C_\ell,~C_r$ and $C_d$ in the experiment, Fig. \ref{AB}(b)) by
connecting each site on the three ring segments to a 1D lead,
which allows only an outgoing current to an absorbing reservoir
(Fig. \ref{AB}(c)). Each such segment is thus replaced by a
``comb" of absorbing ``teeth".

\begin{figure}
\begin{tabular}{cc}
{\epsfxsize=2in\epsfbox{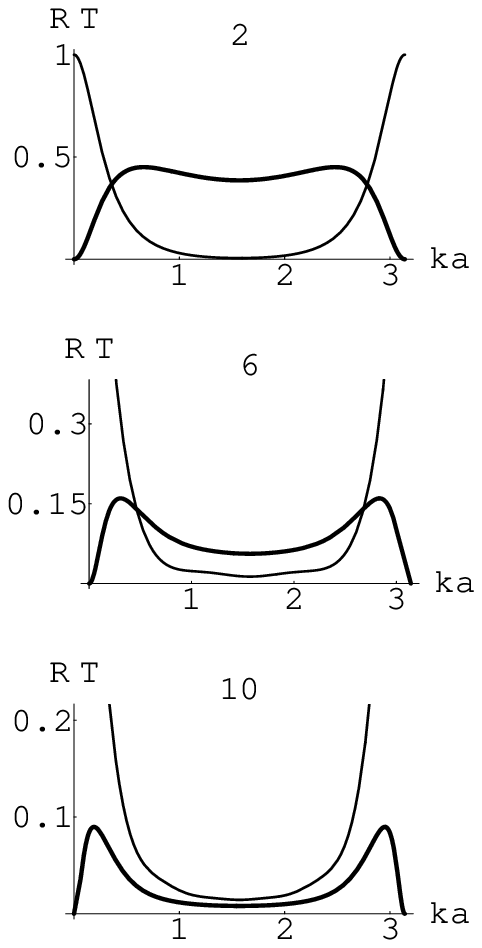}}&
{\epsfxsize=2in\epsfbox{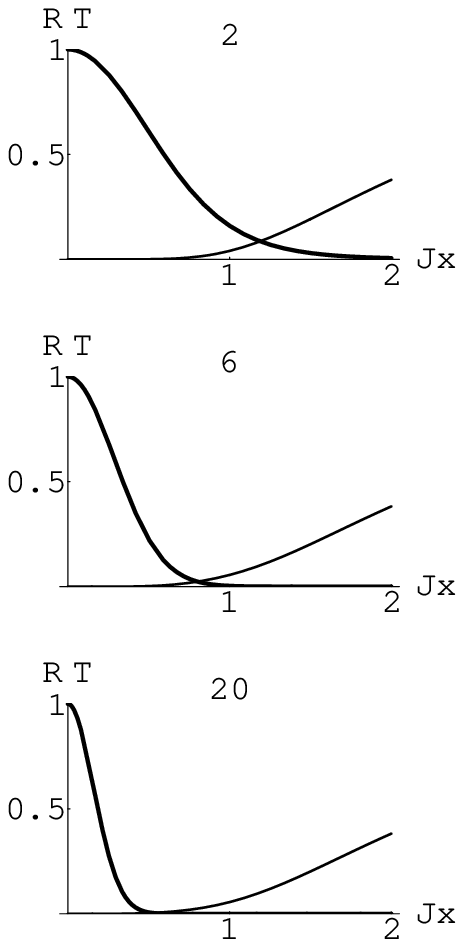}}

\end{tabular}
\vskip .4truecm \caption{Transmission (thick line) and reflection
(thin line) through a ``comb", versus $ka$ at $J_x=.7J$ (left) and
versus $J_x$ at $ka=\pi/2$ (right). The number on each frame gives
the number of ``teeth", $M$.} \label{comb}
\end{figure}

We start by investigating the properties of a single ``comb". The
``base" of the ``comb" is described by a chain of $M$
tight-binding sites, with $J_{m,m+1}=J_c$ and $\epsilon_m=0$. Each
``tooth" is represented by a 1D tight-binding chain, with
$\epsilon_j=0$. The first bond on the ``tooth" has $J_{m,0}=J_x$,
while $J_{j,j+1}=J$ for $j \ge 0$. Assuming only outgoing waves on
the teeth, with wave functions $t_x e^{ikaj}$ and energy $E=-2J
\cos ka$, one can eliminate the ``teeth" from the equations. The
wave functions on the ``base" of the comb are then given by
$\psi^c_m=A_c \eta_c^m+B_c \eta_c^{-m}$, where $\eta_c$ is a
solution of the (complex energy) equation $E+J_x^2
e^{ika}/J=-J_c(\eta_c+\eta_c^{-1})$. When this ``comb" is treated
as our basic scatterer, i.e. connected via $-J_{in}$ and
$-J_{out}$ to our ``standard" two leads, then the transmission and
reflection amplitudes via the ``comb" are given (up to unimportant
phases) by $t=J_{out}(A_c \eta_c^N+B_c/\eta_c^N)/J$ and
$r=J_{in}(A_c \eta_c+B_c/\eta_c)/J-e^{ika}$, and one ends up with
two linear equations for $A_c$ and $B_c$. The results for
$T=|t|^2$ and $R=|r|^2$ are shown, for three values of $M$, in
Fig. \ref{comb}, as functions of $ka \in [0,\pi]$ in the free
electron energy band, for $J_x=.7J$ (left), and as functions of
$J_x$, for $ka=\pi/2$ (right). In the figure,
$J_c=J_{in}=J_{out}=J$. It is rewarding to observe that both $T$
and $R$ are almost independent of the electron energy $E$ over a
broad range near the band center. It is also interesting to note
that for these parameters, $T$ decreases with $J_x$, but $R$
increases with $J_x$. For fixed $J_x$, $T$ and $R$ exhibit some
even-odd oscillations with $M$, but basically $T$ decreases with
$M$ while $R$ increases towards an almost constant value for
$M>6$. This is understandable: a strong coupling to the ``teeth"
causes a strong decay of the wave function along the ``comb".
Thus, for each value of $M$ one can find an intermediate optimal
region in which both $T$ and $R$ are small. This region broadens,
and has smaller $T$ and $R$, for larger $M$.

We next place three such ``combs" on the AB interferometer, as in
Fig. \ref{AB}(c), and study the AB transmission $T$ as function of
the various parameters. For simplicity, we set the same parameters
for all the combs, and vary the coupling strength $J_x$. Since
each ``tooth" of the ``comb" can be replaced by adding the complex
number $J_x^2e^{ika}/J$ to the energy $E$ in the equations for
$\psi^s_m$ on the ring segments, the mathematics is similar of
that of the ``bare" closed interferometer. The main difference in
the results is that now $\eta_c$ is complex, yielding a decay of
the wave function through each comb. This also turns the ratio
$F/G$ in Eq. (\ref{tt}) complex, yielding non-trivial values for
$\beta$. To demonstrate qualitative results, we choose
$M_\ell=M_r=6,~M_d=12$, use $J_\ell=J_r=J_d=J_c=J$ and keep
$ka=\pi/2$ and the QD parameters $J_L(n)=J_R(n)=J,~N=4,~U=20J$.
The choice for the ``comb" parameters ensures that $A$ and $B$ in
Eq. (\ref{fit}) are of the same order. Other choices give similar
qualitative results. Fig. \ref{ziur} shows results for $A,~B,~C$
and $\beta$ as function of $V$, for several values of $J_x$.
Clearly, $J_x=.15J$ gives a phase $\beta$ which is intermediate
between the Onsager jumps of the upper left Fig. \ref{ziur} and
the exact intrinsic $\alpha_{QD}$ of Fig. \ref{alpha}. Increasing
$J_x$ yields a saturation of $\beta$ onto $\alpha_{QD}$, which
persists for a broad range between $J_x=.5J$ and $J_x=.9J$.
However, larger values of $J_x$, e.g. $J_x=1.5J$, cause a
deviation of $\beta$ from $\alpha_{QD}$, due to the increase of
the reflection from each ``comb". Interestingly, this deviation is
{\bf in the same direction} as for small $J_x$! The reason for
this is clear: as the reflection of each comb increases, the
electron ``rattles" in and out of the QD. This localizes it on the
QD, and reduces the width of the QD resonances. For these large
values of $J_x$, one has $|Z/W| \ll 1$ in Eq. (\ref{tt}). Thus,
the two-slit conditions hold, and one has $B \propto |t_1|$ and
$\beta=\alpha_1$. We have solved the equations for the
transmission through the upper branch only (disconnecting the
lower branch altogether), and found that indeed, the coefficient
$c$ in $t_1=c t_{QD}$ is a constant as long as the reflection of
the combs is small. However, as $J_x$ increases above about $.9J$,
$c$ is no longer a constant. The narrower resonances shown in Fig.
\ref{ziur} (lower right) fully agree with this modified upper
branch transmission. In any case, ``optimal combs", with small $T$
{\it and} $R$, do yield $\beta=\alpha_{QD}$.

So far, we assumed {\it no} direct losses from the QD itself. It
is easy to add such losses, by connecting a ``lossy" channel to
each resonant state $n$ \cite{prl}, similar to the ``teeth" of our
``combs", with a tunneling amplitude $J_x'$. As before, this
introduces a complex addition $J_x'^2e^{ika}$ to $E-E_R(n)$. Fig.
\ref{loss} shows the results for the same parameters as above, but
with $J_x=J_x'=.9J$. Clearly, the new imaginary parts eliminate
the Fano-like zero in $B$, and yield a smooth variation of $\beta$
near the ``intrinsic phase lapses". Although similar to the
behavior arising in the BW approximation, the present effects are
{\it real}, due to physical breaking of the unitarity on the QD.
It is interesting to note that the data of Ref. \cite{schuster}
show similar (and otherwise unexplained) smooth features. It is
however possible that the latter come from finite temperature
averaging\cite{sun}.

\section{Discussion}

In conclusion, we find that the AB interferometer yields
quantitative information on the QD resonances only if the electron
crosses each segment on the ring, as well as the QD itself, only
once. As stated, this can be rephrased by two criteria: having the
two-slit condition -- i.e. effectively no reflections back from
the ``forks" into the ring's branches, and having no ``rattling"
around the QD -- i.e. having little reflection from the ``lossy"
terminals. A third criterion requires no direct losses from the QD
itself.

\begin{figure}\begin{tabular}{cc}
\epsfbox{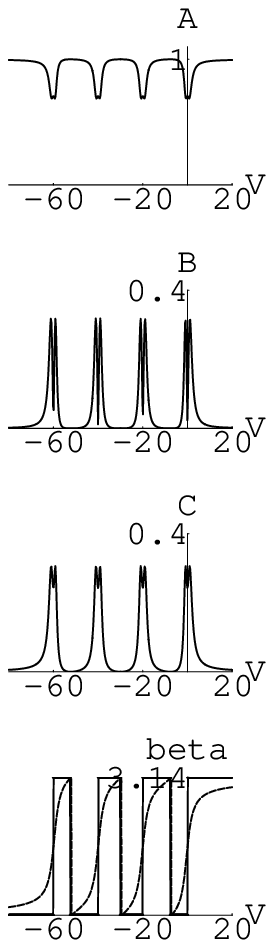}& \epsfbox{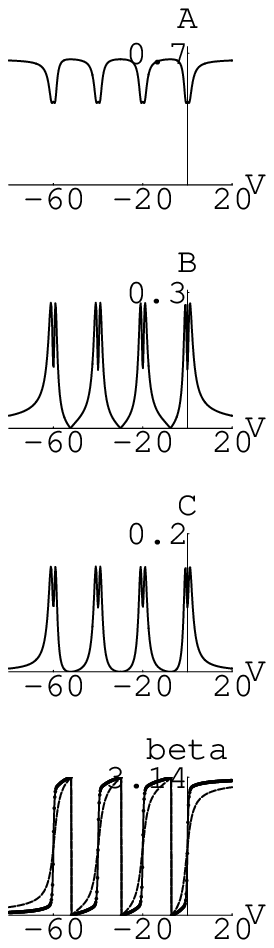}\\
\epsfbox{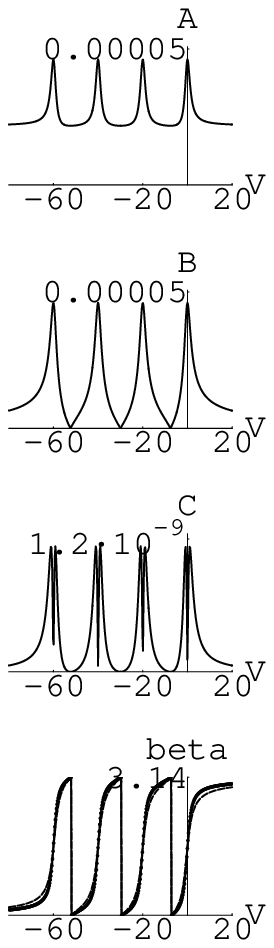}&\epsfbox{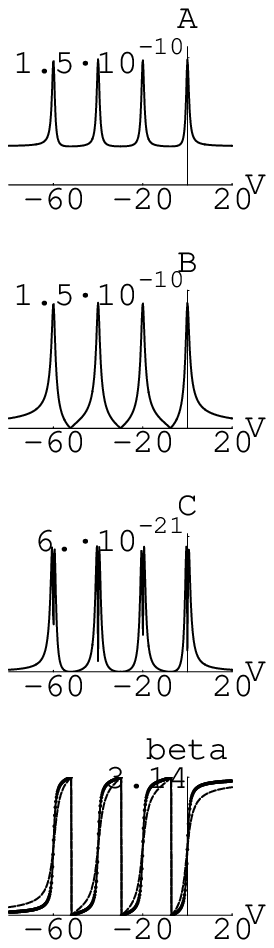}

\end{tabular}

\vskip .4truecm \caption{$A,~B,~C$ and $\beta$ for transmission
through the closed AB ring (upper left), and for the open
interferometer with $J_x=.15J$ (upper right) and $J_x=.9J,~1.5J$
(lower left, right). The dashed line shows the exact intrinsic
phase $\alpha_{QD}$.} \label{ziur}
\end{figure}

\begin{figure}
\epsffile{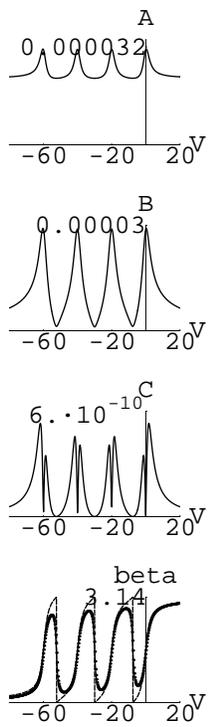}
\vskip .4truecm \caption{Same as Fig. \ref{ziur}, but with a
``lossy" channel attached to the QD; $J_x=J_x'=.9J$.} \label{loss}
\end{figure}

The two-slit conditions are easy to examine: a small $|Z/W|$ in
Eq. (\ref{tt}) implies small amplitudes for all except the first
harmonic in Eq. (\ref{fit}), as indeed seen by the decreasing
relative values of $C$ for increasing $J_x$ in Fig. \ref{ziur}.
This is also easily checked in the analysis of the experimental
data\cite{schuster}. It might be interesting to fit intermediate
range data to the exact Eq. (\ref{TT}), instead of using a
truncated Fourier series as in Eq. (\ref{fit}).

The second condition, which has not been emphasized in the
literature before, is somewhat harder to confirm. One way to check
this is to vary $J_x$ experimentally, and look for the value which
gives the {\it largest} width of the resonances. Other ways
require disconnecting the lower branch, and studying the
conductance through the ``lossy" path including the QD and the two
``combs". The ``combs" are acceptable for our purposes only below
a threshold $J_x$, as long as the conductance peaks remain
independent of $J_x$.

It is worth emphasizing that the experimental data (as reflected
in Fig. \ref{ziur}) actually contain more than the AB phase shift
$\beta$. As stated after Eq. (\ref{2slit}), the two-slit condition
implies that $B=2|t_1t_2|$. Since $t_2$ is independent of $V$,
this gives $B \propto |t_1|$. Assuming also that the ``combs" on
the upper branch do not modify the $V$ dependence (i.e. that $c$
is $V$-independent), we conclude that
$T_{QD}=|t_{QD}|^2=(B/B_{max})^2$, where $B_{max}$ is the maximum
of $B(V)$.
 Indeed, we confirmed that our
``data" in Fig. \ref{ziur} obey this relation in the optimal range
of $J_x$. Moving away from these optimal conditions causes a
steeper increase in $\beta$, and a related narrower peak in $B$.
Both of these widths should be largest for the optimal conditions.
In fact, a third way to ensure a correct measurement of
$\alpha_{QD}$ would be to measure $T_{QD}$ directly, from the
conductance of the {\it isolated} QD, and compare it with the
normalized $B^2$ in the interferometer measurement. Obviously, all
of the latter experiments require modifications of the mesoscopic
circuitry, and may thus not be straightforward to follow.

Although we presented results for only one set of parameters, we
emphasize that similar results can be obtained for many other
sets. In particular, the results for $\beta$ and for $B/B_{max}$
do not depend on the parameters of the lower branch. Varying these
parameters only adds $V$-independent factors, and changes the
$V$-dependence of $A$ (which is dominated by the ratio
$|t_1/t_2|$). The results are also not sensitive to the sizes
$M_s$ of the ``combs". However, too broad combs imply too small
values of the total transmission through the interferometer (at
optimum), giving very small outgoing currents which may be
difficult to measure. Thus, although Weidenm\"uller\cite{weiden}
is right in wishing many terminals, this is not enough. One could
also vary other parameters, like $J_c$, but this might introduce
additional resonances, due to the ``combs" and not to the QD.
Similar undesired comb-related resonances also arise when $ka$ is
close to the band edge, but will not arise when one abandons the
special 1D treatment of the leads and branches, a situation which
is better modeled near the center of the band.

Our analysis also shows that even away from optimum, the {\it
locations} of both the resonances and the Fano-like zeroes (or
``phase lapses")  are reproduced correctly, independently of the
coupling strength $J_x$. The main purpose of optimizing the
interferometer is thus to obtain accurate values of the intrinsic
resonance {\it widths}, which should agree with those found from
the direct measurements of the peaks in the isolated QD
conductance.

\acknowledgements

We thank M. Heiblum, Y. Levinson, A. Schiller, H. A.
Weidenm\"uller and A. Yacoby for helpful conversations. This
project was carried out in a center of excellence supported by the
Israel Science Foundation, with additional support from the Albert
Einstein Minerva Center for Theoretical Physics at the Weizmann
Institute of Science, from the German Federal Ministry of
Education and Research (BMBF) within the Framework of the
German-Israeli Project Cooperation (DIP) and from the NSF grant
DMR 99-81283.

\newpage

\end{multicols}

\begin{references}
\bibitem{note} For the purposes of the present paper, it does not matter if the
resonant state on the QD involves electron-electron interactions.
\bibitem{T0} We assume zero temperature, so that $E$ is equal to
the Fermi energy. Finite temperature replaces the sharp ``phase
lapses" by smooth but fast changes \cite{sun}.
\end{references}
\end{document}